\documentclass[superscriptaddress,twocolumn,preprintnumbers,amsmath,amssymb,aps]{revtex4}

\pdfoutput=1
\usepackage[utf8]{inputenc}
\usepackage[USenglish]{babel}

\usepackage{amsmath,amssymb,amsthm,amsfonts}
\usepackage{MnSymbol}
\usepackage{mathrsfs}
\usepackage{slashed}
\usepackage{graphicx}
\usepackage{subfigure}
\usepackage{color,rotating}
\usepackage{dsfont}
\usepackage{setspace}
\usepackage{verbatim}
\usepackage{hyperref,doi}
\usepackage[export]{adjustbox}
\usepackage[normalem]{ulem}
\usepackage{bbold}

\usepackage[sort&compress]{natbib}

\makeatletter
\def\l@subsubsection#1#2{}
\makeatother

\newcommand{\beq}{\begin{equation}}
\newcommand{\eeq}{\end{equation}}
\newcommand{\beqa}{\begin{eqnarray}}
\newcommand{\eeqa}{\end{eqnarray}}
\newcommand{\bfc}{\begin{figure}[t]\begin{center}}
\newcommand{\efc}{\end{center}\end{figure}}

\def\fig#1{Fig.~\ref{#1}}
\def\tab#1{Table~\ref{#1}}

\def\app#1{Appendix~\ref{#1}}

\def\0#1#2{\frac{#1}{#2}}  

\graphicspath{{./figures/}}

\newcommand{\diag}{\mathrm{diag}}

\newcommand{\be}{\begin{eqnarray}}
\newcommand{\ee}{\end{eqnarray}}

\begin{document}
\title{Discovering Symmetry Invariants and Conserved Quantities\\by Interpreting Siamese Neural Networks}

\author{Sebastian J. Wetzel}
\affiliation{Perimeter Institute for Theoretical Physics, Waterloo, Ontario N2L 2Y5, Canada}
\author{Roger G. Melko} 
\affiliation{Perimeter Institute for Theoretical Physics, Waterloo, Ontario N2L 2Y5, Canada}
\affiliation{Department of Physics and Astronomy, University of Waterloo, Waterloo, Ontario, N2L 3G1, Canada}
\author{Joseph Scott}
\affiliation{School of Computer Science, University of Waterloo, Waterloo, Ontario, N2L 3G1, Canada}
\author{Maysum Panju}
\affiliation{Department of Statistics and Actuarial Science, University of Waterloo, Waterloo, Ontario, N2L 3G1, Canada}
\author{Vijay Ganesh} 
\affiliation{Department of Electrical and Computer Engineering, University of Waterloo, Ontario, Canada}

\begin{abstract}
We introduce interpretable Siamese Neural Networks (SNN) for similarity detection to the field of theoretical physics. More precisely, we apply SNNs to events in special relativity, the transformation of electromagnetic fields, and the motion of particles in a central potential. In these examples, SNNs learn to identify datapoints belonging to the same event, field configuration, or trajectory of motion. We demonstrate that in the process of learning which datapoints belong to the same event or field configuration, these SNNs also learn the relevant symmetry invariants and conserved quantities. Such SNNs are highly interpretable, which enables us to reveal the symmetry invariants and conserved quantities without prior knowledge.
\end{abstract}

\maketitle


\section{Introduction}
Machine learning (ML) algorithms have experienced a surge in the physical sciences. This is based on the introduction of ML methods to fulfill tasks beyond the scope for which they were originally designed. These include finding phase transitions \cite{Carrasquilla2017,Nieuwenburg2017,Wang2016,Wetzel2017,Zhang2017,Schindler2017,Hu2017,Ohtsuki2017,Broecker2017,Deng2017,Chng2017,Huembeli2018}, simulating quantum systems \cite{Torlai2016,Carleo2017,Inack2018,Hibat-Allah2020,Carrasquilla2019,Ferrari2019,Sharir2020} and rediscovering physical concepts \cite{Schmidt2009,Iten2020,Wetzel2017a,Ponte2017,Greitemann2019,Mototake2019,Udrescu2019,Wang2019}.

Even though ML in theoretical physics is a young discipline, it has so far been successful in reproducing results in many complicated systems in just a few years. This success often comes at the cost of a lack of understanding of what ML algorithms intrinsically learn. Physics, as a scientific discipline, benefits from a ``deeper understanding'' of the underlying models used for making predictions. 

The question of whether ML models can ``understand'' physics is a deeply philosophical one, and we don't presume to address it in all its complexity. Assuming that a ML algorithm is successfully trained to predict the outcome of a physical experiment or calculation, it is not always clear whether the algorithm has deduced physical concepts or has merely managed to perform some basic pattern matching. However, if the ML model is ``interpretable'' in the sense that we can recover a compact and simple mathematical representation of a physical equation by analyzing the said model, then we take the position that such a model has indeed learned to ``understand'' the underlying physics.

The most successful ML algorithms are artificial neural networks (ANNs), which are famously inscrutable. Having said that, there have been many recent attempts at interpreting the learned features of a fully trained ANN. The simplest way to interpret a neural network is to examine the weights and biases of individual neurons, which can only yield successful results in shallow ANNs. In the field of explainable AI (xAI), there are different methods that determine which features of the given input are responsible for a learned model's classification \cite{NIPS2017_7062,gunning2017explainable}. Similarly, in the field of computer vision, there have been many developments to examine the contribution of the pixels in an image to the ANN prediction \cite{Montavon2018,Ribeiro2016,Simonyan2014,Bach2015,Zeiler2014}. One of the most popular methods is feature visualization by enhancing learned patterns on input images \cite{Mordvintsev2015}.

In physics, one has a distinct advantage when it comes to interpreting ANNs. In the field of computer vision or natural language processing, it is very hard to come up with mathematical equations uniquely describing the ``ground truth''. In contrast, physicists have worked for hundreds of years to formulate their theories and experimental measurements in terms of mathematical equations. This means that if we can recover such an equation by analyzing an ANN, we have an immediate access to its interpretation. This also opens up the possibility to check for consistency and reveal new concepts. Indeed a few recent works have presented successful interpretations of ANNs in physics \cite{Wetzel2017a,Kim2018,Suchsland2018,Zhang2019,Bluecher2020}. 

In this article, we propose a change in traditional Siamese Neural Networks (SNN) architectures that makes them easier to interpret. Specifically, the key feature is a {\it bottleneck} layer, where the SNN is forced to compress all available information from previous layers. The output of this bottleneck can be analyzed, for example, by applying known regression methods. A similar approach has been taken in \cite{Wetzel2017a}. While there does not exist a related interpretation procedure in computer vision, the idea of interpreting bottleneck layers is also seen in disentangling autoencoders \cite{Burgess2018}.

The ANNs we are considering in this work are a variant of the previously proposed SNNs, a class of ANNs that have been applied to object tracking, face recognition, and image similarity detection~\cite{Bromley1993,Chopra2005,Taigman2014,Appalaraju2017}. An SNN consists of two (identical) ANNs that are applied to a pair of input data points. The two networks share their weights and biases, which are updated simultaneously during training. The goal of the network is to map the input pairs to a set of latent variables that determine the similarity of the pair. 

The general problem an SNN attempts to solve can be stated as follows: given two data points $x$ and $y$ related by an equivalence relation (e.g., the same event in a relativistic setting measured by two observers in different reference frames), is it possible to correctly and automatically classify them as ``related''? Further, if $x$ and $y$ are not related, then we require the ANN to classify them as not related. 

SNNs can solve an extension of a classification problem with relatively little training data per class. Instead of training a traditional neural network to distinguish between a fixed number of classes, an SNN can probe the similarity of one datapoint with another prototypical datapoint for a certain class. This reformulation bears many advantages. First, the number of classes does not need to be fixed. Further, it is no longer necessary to train on all of the classes. A successfully trained SNN might be able to share its learned representation to distinguish between classes that are not in the training set. These properties become important in the limit of many (possibly infinitely many) classes, or in the case where only a few data points are available in each class.

\subsubsection*{Contributions}

The contributions we make in this paper are:

\begin{enumerate}
\item We introduce the SNN to the field of theoretical physics.

\item We demonstrate its usage in the well known contexts of special relativity, electromagnetism, and the motion of particles in a central potential. In the case of special relativity, these SNNs learn whether or not two different observations of physical phenomena correspond to the same event. In the case of electromagnetism, these SNNs learn whether or not given two field configurations, one can be transformed into the other via a Lorentz transformation. In the case of motion of particles, these SNNs discover whether or not two observations of position and momenta describe the same particle. 

\item Further, we successfully interpret the intermediate output representations of the SNN and recover the mathematical formulations of known physical conserved quantities and invariants, e.g., the spacetime interval or the angular momentum.

\item Since the interpretation of the SNN yields signatures of known physical equations, we argue that our SNN has indeed learned to ``understand'' physical concepts instead of merely performing basic pattern matching.
\end{enumerate}

\newpage
\section{Neural Network Architecture}

\begin{center}
\begin{figure}[htb!]
\includegraphics[width=0.5\textwidth]{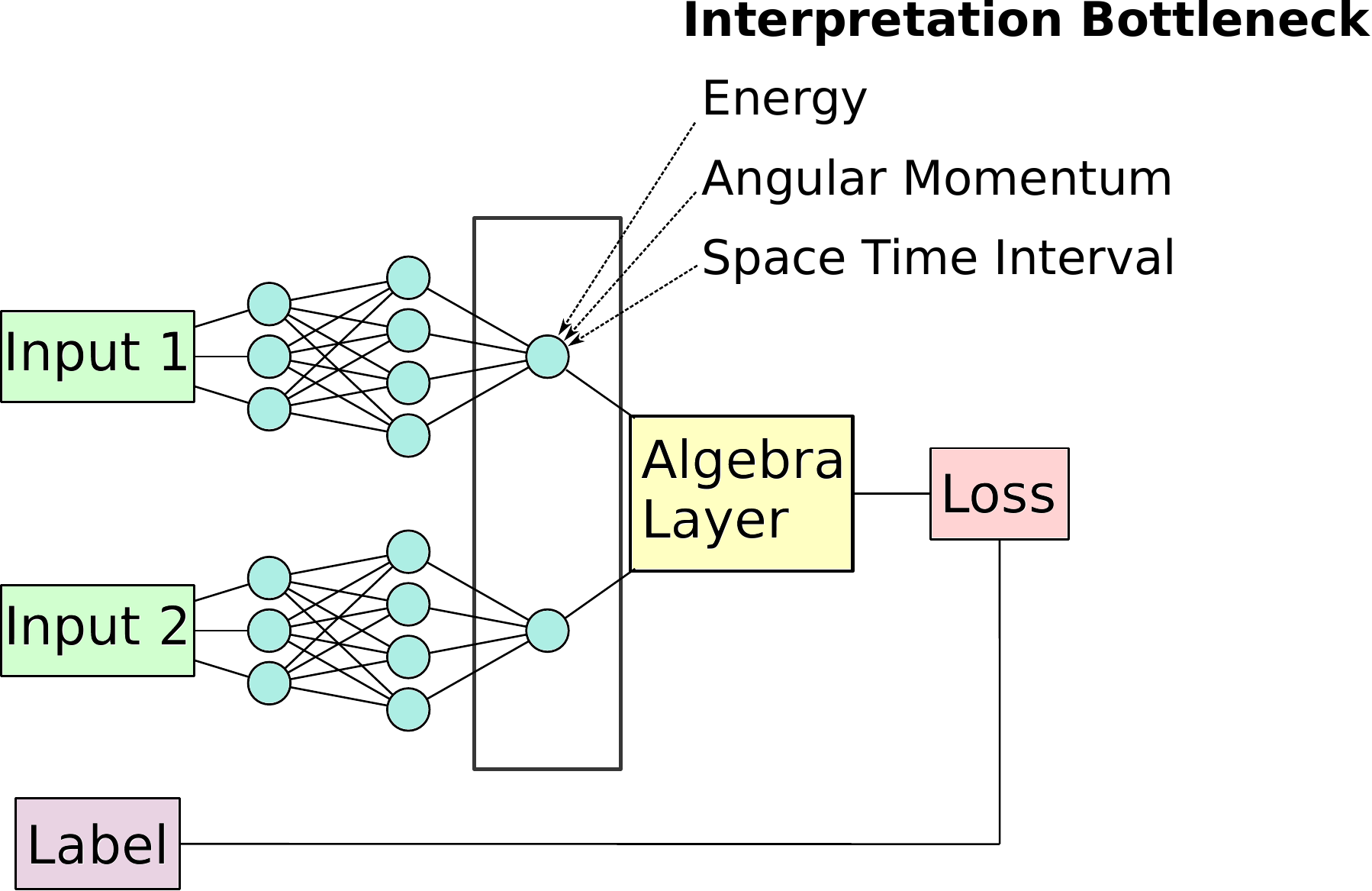}\\
\caption{Schematic architecture of an interpretable SNN. Our SNN contains a bottleneck of only a single neuron, the output of this layer is called the intermediate output of the network. We observe that this bottleneck encodes quantities which are strongly correlated with invariants like the energy or the spacetime interval.}
\label{fig:architecture01}
\end{figure}
\end{center}

In this paper, we employ Siamese Neural Networks (SNN) to determine whether or not two samples belong to the same class. In this context, our input data is a pair of samples $X_i=(x_i,x_i^\prime)$. In order to formulate a supervised learning problem, we associate the label $y_i=0$ to pairs that correspond to the same class (i.e., $x_i$ and $x_i^\prime$ are related via an equivalence relation) and $y_i=1$ to pairs that belong to different classes (i.e., the input pairs are not related). In this sense, we can reformulate a classification problem with many -- possibly infinitely many -- classes into a traditional binary classification problem.

For this purpose, we construct our SNN consisting of several building blocks. The first building block is composed of a pair of identical neural networks. This pair of networks is applied simultaneously to each of the samples in a data point pair $x_i$ and $x_i^\prime$. The last layer of the network pair only contains a single neuron, this layer we refer to as the bottleneck. The output of the bottleneck layer is the intermediate output of the SNN. The intermediate output is merged by performing appropriate algebraic operations. Let us denote $f(x_i)$ and $f(x_i^\prime)$ the output of each of the neural networks. Then the algebra layer calculates $(f(x_i)-f(x_i^\prime))^2$ before supplying it to a sigmoid neuron such that the output of the full SNN can be written as
\begin{align}
    F(X_i)=\mathrm{sigmoid} \left(  w \, (f(\mathbf{x}_i)-f(\mathbf{x}_i^\prime))^2 \, + \, b    \right)\quad . \label{eq01}
\end{align}


\begin{center}
\begin{figure*}
\includegraphics[width=1.0\textwidth]{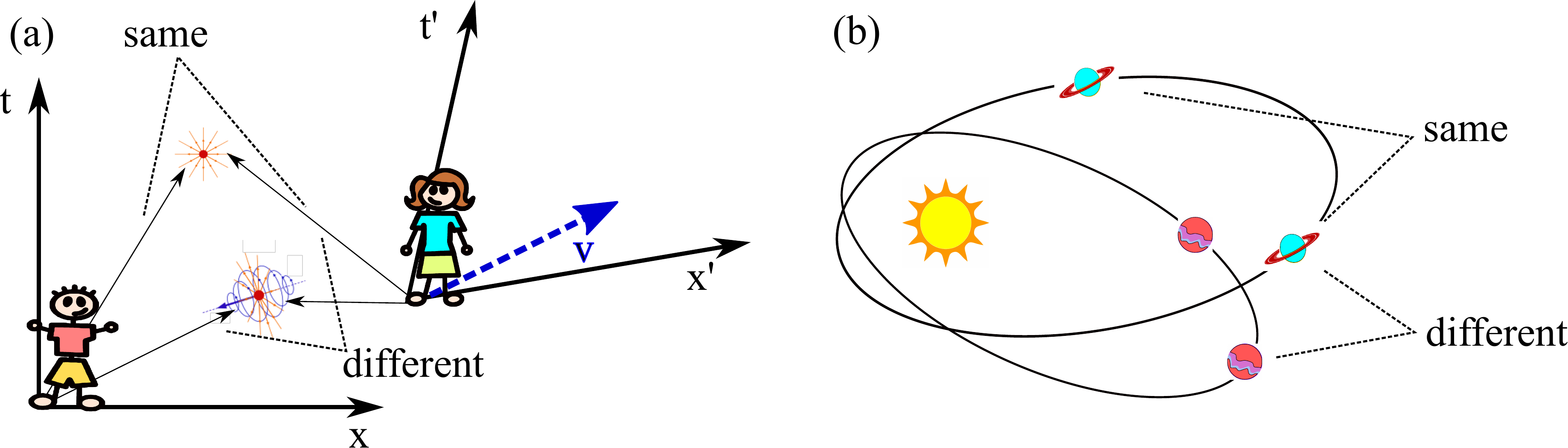}
\caption{Schematic description of the task solved by the Siamese neural network. Left (a): In the case of special relativity and electromagnetism our network is tasked to learn if two descriptions from different perspectives correspond to the same event or the same electromagnetic field configuration. Right (b): In the context of Newtonian gravity, we train our network to detect if two observations of velocities and positions correspond to the same particle moving in a central potential.}
\label{fig:stickfig}
\end{figure*}
\end{center}


The SNN outputs a probability that signifies whether the two samples belong to the same class or not. 

For the purpose of training, we minimize the binary cross-entropy loss function between the SNN $F(X_i)$ and the label $y_i$ on the training set. After training is complete, the generalization performance is measured on the test set. While training the SNN, we enforce weight sharing in the network pair to make sure these networks learn the same function. We note that, in the context of our physical examples, a natural minimization of the binary cross-entropy loss function is achieved if $f(x_i)$ learns a symmetry invariant or a conserved quantity.

After having successfully trained the SNN, our goal is to answer the question of which features this neural network bases its decision. In general, there is no easy answer to this question, since analyzing even small neural networks can be extremely challenging. So far, there does not exist a comprehensive theory of what is learned by artificial neural networks.

One of our crucial insights is that in order to interpret what our SNN learns, we have designed the SNN to include a bottleneck at the output of the first building block before merging (see \fig{fig:architecture01}). We will see later that our SNN learns conserved quantities and invariants at the bottleneck in order to make its decision about whether two samples belong to the same class. Further, by interpreting the network, we can predict conserved quantities and invariants with no additional prior knowledge.

If the number of neurons in the bottleneck layer increases, one can achieve better accuracy at the cost of interpretability. The interpretability can, in principle, be retained if one enforces decorrelated intermediate outputs.

More details about our neural network architecture and the learning procedure can be found in \app{app:nn}.

\section{Performing Machine Learning}

\subsection{Spacetime in Special Relativity}
\subsubsection*{Introduction}
The first physical system we consider in this work is the Minkowski spacetime in special relativity. An event is a four vector $(t,x,y,z)\,  \in \mathbb{R}^4 $ that combines spatial coordinates and a moment in time. Minkowki spacetime is $\mathbb{R}^4$ with a scalar product induced by the metric $\eta_{\mu \nu}=\diag(-1,1,1,1)$,
\begin{align}
    \langle \mathbf{x},\mathbf{y}  \rangle&= \eta_{\mu \nu} x^\mu y^{\nu} = x_\mu y^{\nu} \quad,
\end{align}
where we have used $x_\mu=\eta_{\mu \nu}x^\nu$.  Thereby we define the spacetime interval $s$ by
\begin{align}
    \langle \mathbf{x},\mathbf{x}  \rangle &=-t^2+x^2+y^2+z^2=s^2 \quad .
\end{align}
The Lorentz Group is the set of transformations which preserve the scalar product on Minkowksi spacetime
\begin{align}
  O(3,1)=\left\{ \Lambda \in \mathcal{M}(\mathbb{R^{4}}): \langle \Lambda \mathbf{x},\Lambda \mathbf{y}\rangle=
  \langle \mathbf{x},\mathbf{y} \rangle \,  \forall \, \mathbf{x},\mathbf{y}  \in \mathbb{R}^{4}\right\}
\end{align}
and thus also preserve the spacetime interval.
\subsubsection*{SNN Training}
\begin{center}
\begin{figure}[htb!]
\includegraphics[width=0.5\textwidth]{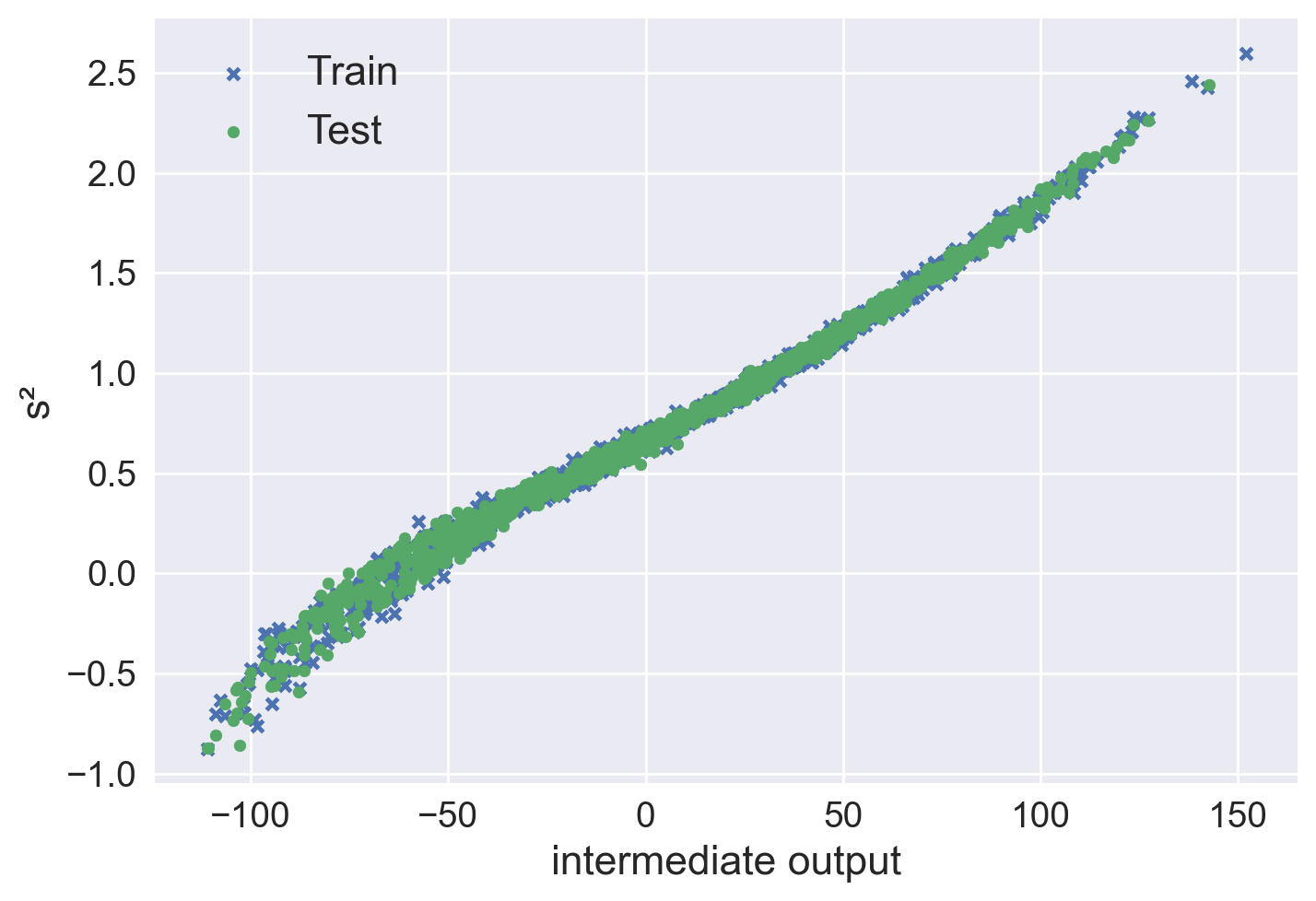}\\
\caption{Special Relativity: Correlation between the intermediate output of the siamese neural network at the bottleneck layer and the spacetime interval.}
\label{fig:sr_cor1}
\end{figure}
\end{center}

In this section, we discuss how to teach the neural network to identify, in special relativity, whether two observations by different observers correspond to the same event. These observers are at the same position but move with a relative velocity in some direction. For this purpose, we prepare positive training data of pairs of observations that correspond to the same event, and negative data where a pair of measurements does not describe the same event.

More specifically, in order to train our neural networks with data we prepare a training dataset consisting of pairs of measurements of the same event in Minkowski spacetime seen from two different observers  $X^\mu=(\mathbf{x}^\mu,\mathbf{x}^{\prime\mu}=\Lambda^{\ \mu}_\nu \mathbf{x}^\nu)=((t,x,y,z),(t^\prime,x^\prime,y^\prime,z^\prime))$. Here, $\Lambda $ is a random Lorentz transformation which is sampled from all possible Lorentz transformations. More details can be found in \app{app:lt}. We sample 50000 spacetime events $\mathbf{x}^\mu$ and Lorentz transformations $\Lambda$ to create pairs of events that form the positive dataset. We associate with each pair the label $y=0$. Further, we create a negative dataset where each pair of spacetime coordinates is not related by a Lorentz transformation. In practice, we implement this by randomly permuting among all second elements of all pairs of spacetime events in the positive dataset. Each pair in the negative dataset is labeled with $y=1$. In addition to this training set, we prepare a similar test dataset of 5000 positive pairs and 5000 negative pairs.

The SNN is trained to predict if a pair of observations describe the same event or not. This is done by optimizing the weights of the neural network via backpropagation to minimize the binary cross-entropy loss between network output $y_p$ and true label $y_t$. After training, the neural network is able to correctly predict if a pair of observations belong to the same event with an accuracy of $\approx 94\% $ on the training dataset and $\approx 92\%$ on the test dataset. The training and testing accuracies during training can be seen in \fig{fig:lossacc}.

Following the successful training of our SNN, we want to understand what the neural network has learned. This can be achieved by examining the intermediate output of the neural network, which acts as an interpretable bottleneck. We perform a hierarchy of linear regressions with polynomial features (i.e.~polynomial regression) on the intermediate output with respect to the input. If we assume that the Taylor expansion of the decision function is sufficiently accurate at the decision boundary, we can hope to get insightful results.

We perform ridge regression with polynomial features of the input on the intermediate output of the SNN. We start with polynomials of degree 1 and increase the degree of the polynomial features until the regression becomes accurate. From \tab{srreg} one can immediately infer that the optimal degree of the polynomial features is 2.

\begin{table}
\begin{tabular}{ c c c } 
 \hline
 \hline
 \hphantom{o} order \hphantom{o} &\hphantom{o} train score \hphantom{o}& \hphantom{o}test score \hphantom{o}\\ 
 \hline
 1 & 0.0013 & 0.0005 \\ 
 2 & 0.9894 & 0.9893 \\ 
 3 & 0.9900 & 0.9899 \\ 
 4 & 0.9907 & 0.9906 \\ 
 \hline
 \hline
\end{tabular}
\caption{Regression scores of the regression on the intermediate output in the case of special relativity, measures normalized distance between regression and data. The score metric is known as the coefficient of determination or $R^2$ score. Best possible score is $1$, the score can be negative.}
\label{srreg}
\end{table}

The result of the regression is in an ordered manner is

\begin{align}
f(\mathbf{x})\approx& -87.41 t^2 -60.48 -0.11 x \nonumber \\
&-0.10 y z  + 0.04 t y+ 0.06 z \nonumber\\
&+0.07 y +0.10 t x + 0.12 t z\nonumber \\  
&+ 0.15 x z + 0.21x y + 2.50 t \nonumber \\ 
&+88.10z^2+ 88.61y^2+88.63x^2 \nonumber\\
\approx& 88\underbrace{(-t^2+x^2+y^2+z^2)}_{=s^2}-60
\end{align}

We can see that four nontrivial features dominate all others. If we assume that the regression includes small approximation errors, we can infer that the SNN has learned the invariant quantity $s^2=-t^2+x^2+y^2+z^2$. This quantity is the spacetime interval, a known invariant of the Lorentz group. In cases where the regression does not yield a clear result, one can cross-check the second-order regression result with higher orders of regression, and observe if the dominant features stay the same. Another option is to do the whole training procedure with a different random seed and see what parts of the results keep the same ratio.

To summarize, as long as the ANN is only able to use a single scalar function to decide if two events are the same, it calculates the spacetime interval. If the spacetime interval is the same, the ANN predicts that both coordinates in a pair belong to the same event. While it is often difficult to decide if neural networks learn to ``understand'' physical concepts to make decisions, here we argue that our SNN does so. To confirm our derivation, we draw a scatter plot for a subset of our data points of the intermediate output versus the spacetime interval in \fig{fig:sr_cor1} and observe a nearly perfect non-linear correlation between these two. Note that we have cross-checked the second-order regression result with higher orders of regression, and found that the dominant features stay the same. 

Finally, we examine whether the SNN can also learn a different quantity to decide if two observations from different observers belong to the same event. For this purpose, we again prepare a training and a test dataset, as explained above. However, in the preparation of the dataset, we keep the spacetime interval fixed. We attempt to train the SNN to learn to associate corresponding observations. However, the ANN fails to train in this case. After the best training cycle, the ANN can only predict if two observations belong to the same event with an accuracy of $58\%$ on the training set or $57\%$ on the testing set, which is barely better than random. This fact leads to the conclusion that the SNN is unable find another invariant of the Lorentz group besides the spacetime interval.

The fact that the SNN fails to distinguish observations in this reduced dataset hints that all observations with the same space-time interval can be transformed into each other by a Lorentz transformation. Further, it indicates that there is no other symmetry invariant. Both of these statements are of course known to be true. However, one needs to be careful since the same conclusions could be drawn if the neural network is not powerful enough to learn an underlying invariant.

\subsection{Motion in a Central Potential}
\subsubsection*{Introduction}
As a second system we consider the motion of a particle in a central potential, such as the movement of a planet in the gravitational potential of the sun. Newtonian gravity can be formulated via the Hamiltonian
\begin{align}
H=\frac{\mathbf{p}^2}{2m}-\frac{G m M}{r}    \quad .
\end{align}
Here $\mathbf{p}$ is the momentum, $r=\sqrt{(x^2+y^2)}$ is the distance from the potential centre, $m$ is the mass of the planet, $M$ is the mass of the sun, and $G$ is Newton's constant of gravitation. Given an initial position $\mathbf{x}$ and velocity $\mathbf{v}$ one can calculate the trajectory of motion by solving Hamilton's equations
\begin{align}
    \dot{x} &= \,\partial_{p_{x}} H \quad \dot{p}_x = \,-\partial_{x} H \nonumber \\
    \dot{y} &= \,\partial_{p_{y}} H \quad \dot{p}_y = \,-\partial_{y} H \quad .
\end{align}
There are conserved quantities in this system, the energy $E$ the components of the angular momentum $\mathbf{L}$ and the components of the Laplace-Runge-Lenz vector $\mathbf{A}$. They are related by two equations which effectively reduces the number of scalar conserved quantities to five.

\subsubsection*{SNN Training}
\begin{center}
\begin{figure}[htb!]
\includegraphics[width=0.5\textwidth]{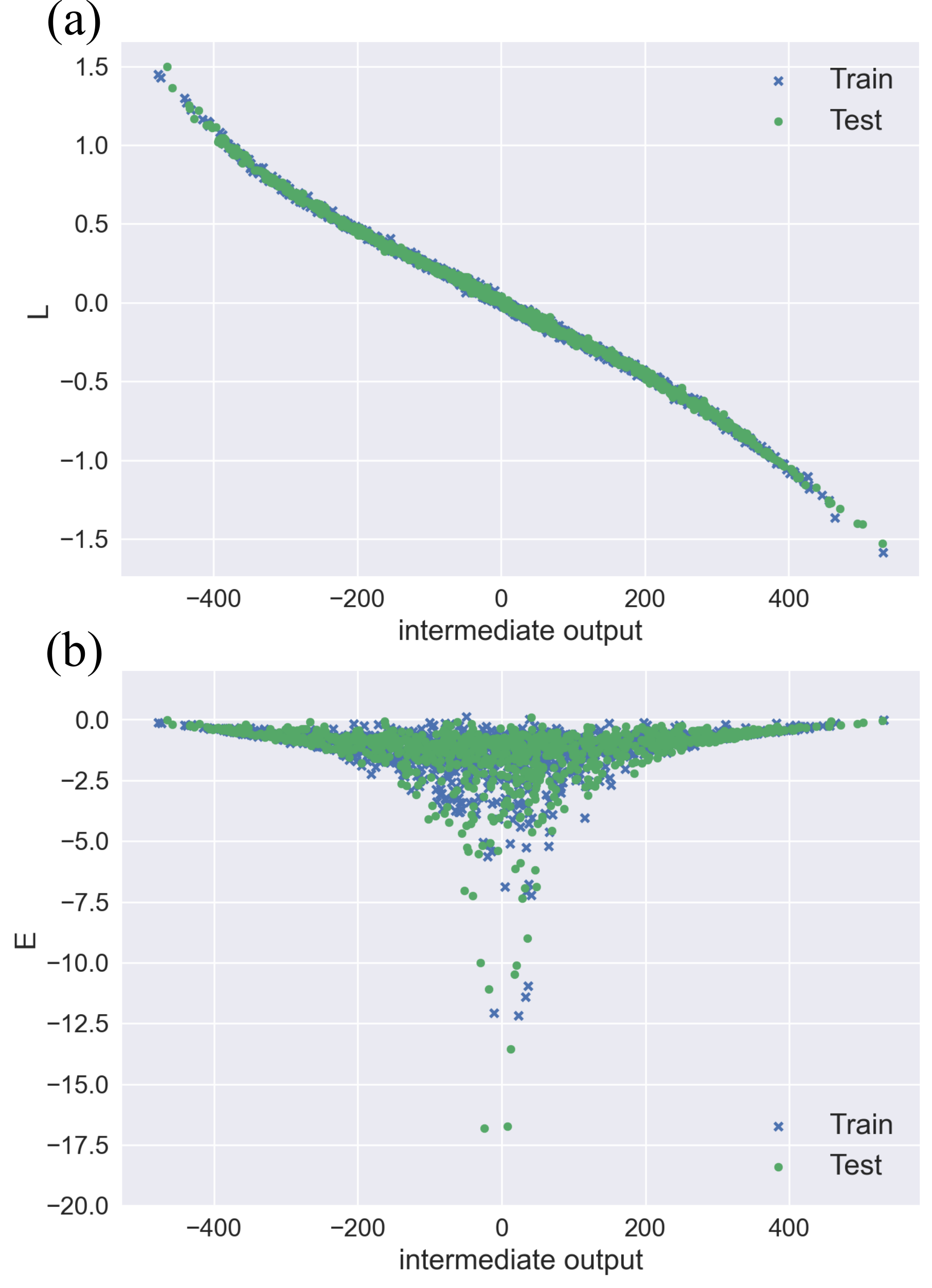}\\
\caption{Particle in central potential: correlation between the intermediate output and (a) the angular momentum or (b) the energy.}
\label{fig:motion1}
\end{figure}
\end{center}
\begin{center}
\begin{figure}[htb!]
\includegraphics[width=0.5\textwidth]{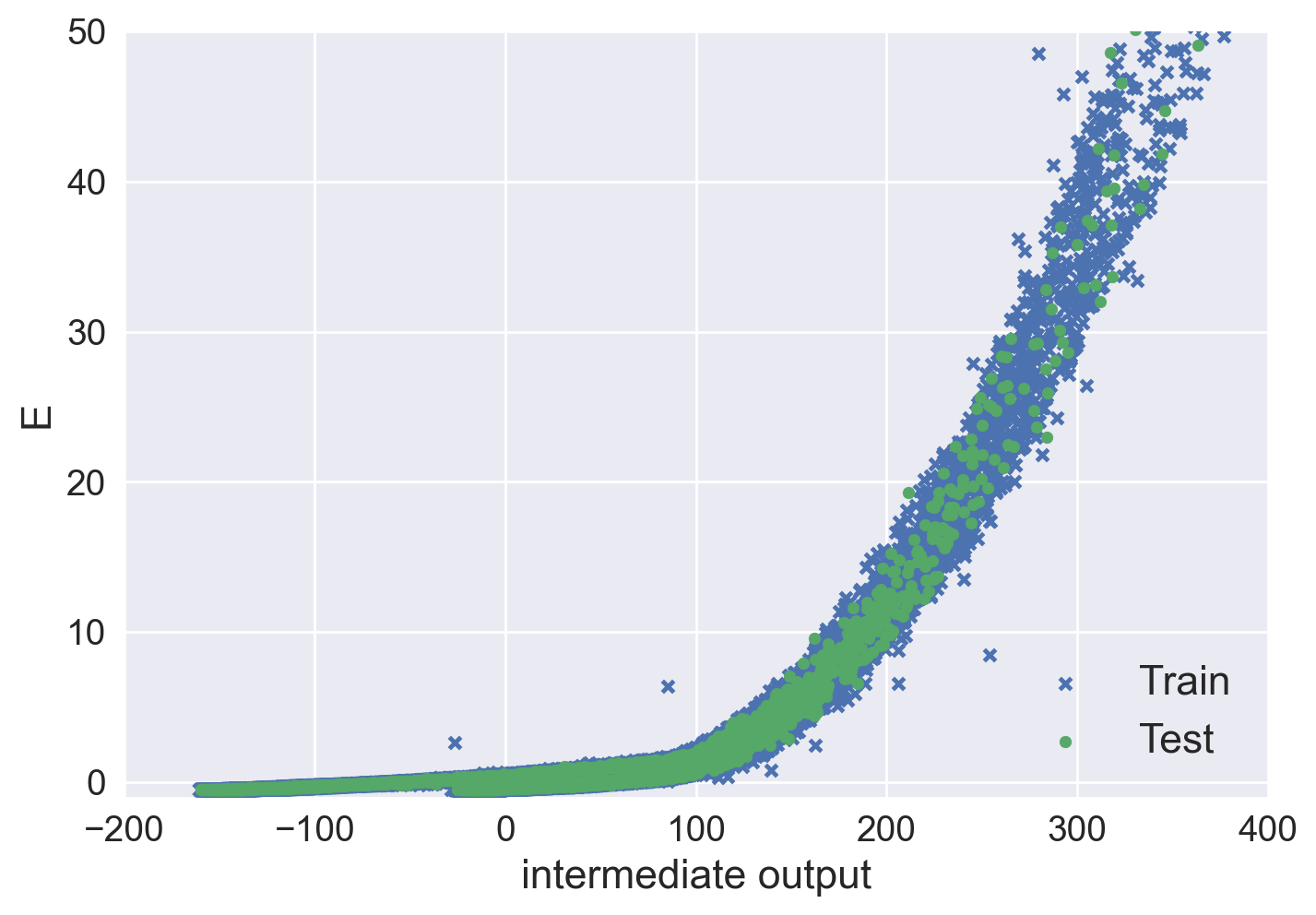}\\
\caption{Particle in central potential with fixed angular momentum: correlation between the intermediate output and the energy.}
\label{fig:motion2}
\end{figure}
\end{center}
When examining the motion of particles in a central potential, the SNN is tasked with determining whether two observations of the same particle correspond to the same particle trajectory.

We simulate particles of fixed mass $m$ moving in a Newtonian static gravitational potential produced by a stationary mass $M$ by solving the Hamiltons equations for a set of random initial positions and velocities. For simplicity we set $m=1$ and $G m M=1$. We measure the position and the velocity of the particle at two different times to get pairs of inputs $X=(\mathbf{x},\mathbf{x}^\prime)=((x,y,v_{x},v_{y}),(x^\prime,y^\prime,v_{x}^\prime,v_{y}^\prime))$. We generate 50000 pairs belonging to the same particle trajectories to form the positive training dataset labelled by $y_i=0$. By permuting the second entry in the pairs, we create a negative dataset labeled with $y_i=1$. Similarly, a testing set is produced with 5000 positive and 5000 negative examples.

The SNN is then trained to correctly predict if a pair of coordinates belong to the same trajectory. After being successfully trained, the network achieves an accuracy of $\approx 98\%$ on the training set and $\approx 97\%$ on the test set. 

In order to interpret on what quantity the neural network bases its decision, we again examine the bottleneck at the intermediate output. We again perform a hierarchy of linear regressions with increasing polynomial features on the intermediate output. The optimal degree of the regression is two (see \tab{motionreg}).
\begin{table}
\begin{tabular}{ c c c } 
 \hline
 \hline
 \hphantom{o} order \hphantom{o} &\hphantom{o} train score \hphantom{o}& \hphantom{o}test score \hphantom{o}\\ 
 \hline
 1 & 0.0003 & -0.0003 \\ 
 2 & 0.9936 & 0.9939 \\ 
 3 & 0.9937 & 0.9940 \\ 
 4 & 0.9952 & 0.9858 \\ 
 \hline
 \hline
\end{tabular}
\caption{Regression scores of the Regression on the intermediate output in the case of the motion of a particle in a central potential.}
\label{motionreg}
\end{table}

The result of the regression in an ordered manner is
\begin{align}
f(\mathbf{x})\approx& -403.71 x v_y -4.85x  -0.58 x y \nonumber\\
&-0.17 x v_x -0.02 v_y^2 -0.01 v_x v_y \nonumber \\
&+ 0.00 v_y^2 + 0.01 v_y + 0.02 v_x \nonumber\\
&+ 0.45 x^2 + 0.66 y^2 + 0.74\nonumber \\
&+ 0.99 y v_y + 1.24 y + 402.44 y v_x \nonumber \\
\approx& -403 \underbrace{(x v_y - y v_x)}_{=L_z}
\end{align}

This quantity is an approximation to the angular momentum $L_x=m(x \, v_y - y \, v_x)$. A confirmation of this result is visualized in the very good correlation between the angular momentum and the intermediate output, illustrated in \fig{fig:motion1} (a). This means that the SNN learns to distinguish between pairs originating from the same trajectory and different trajectories, by calculating the angular momentum. Another conserved quantity in this system is the energy. In \fig{fig:motion1}  (b), we see that the intermediate prediction is not correlated with the energy.

We now fix the angular momentum and perform the simulation again to produce 50000 positive and 50000 negative data pairs. We train the SNN again to distinguish if a pair of observations belong to the same trajectory. Even though the neural network cannot use the angular momentum to determine if the pair corresponds to the same trajectory, it still manages to perform well on this task. The SNN achieves an accuracy of $\approx 99 \%$ on both the training and the test set.

When using linear regression with polynomial features to determine what the SNN has learned in order to make its prediction we fail. On the one hand, there is no clear optimal degree of the polynomial regression. On the other hand, all the regression results do not yield a clear dominant feature. If we compare the intermediate output to the remaining invariants in the system, we find that the intermediate output is strongly correlated with the energy of the system -- see \fig{fig:motion2}. This means that the SNN probes the pair of observations for energy conservation. However, the energy cannot be well approximated by a polynomial function with which we perform our regression.

To circumvent this problem we extend the input features to include the term $1/r=1/\sqrt{x^2+y^2}$ such that the input feature vector reads $(x,y,v_{x},v_{y},1/r)$. From this point we can start the polynomial regression, go on to identify the best polynomial order (see \tab{motionreg2}) and formulate the findings in the following equation
\begin{align}
f(\mathbf{x})\approx& -174.57 -88.28 \frac{1}{r}-87.39 y v_x \nonumber \\
&-1.43 \frac{1}{r^2}+...+1.27 \frac{x}{r}\nonumber \\
&+46.22 v_x^2 +46.53 v_y^2+87.18 x v_y\nonumber \\
&\approx -175+87(\underbrace{x v_y-y v_x}_{=L_z=\text{const}})+90(\underbrace{\frac{1}{2}v_x^2+\frac{1}{2}v_y^2-\frac{1}{r}}_{=E})
\end{align}
We see that the result includes the energy and the angular momentum. The angular momentum evaluates to a constant. Since constants can be absorbed, we conclude that the SNN has learned energy conservation. One might ask whether the SNN is able to find the Laplace-Runge-Lenz vector, which remains open for further investigation.

\begin{table}
\begin{tabular}{ c c c } 
 \hline
 \hline
 \hphantom{o} order \hphantom{o} &\hphantom{o} train score \hphantom{o}& \hphantom{o}test score \hphantom{o}\\ 
 \hline
 1 & 0.0019 & 0.0069 \\ 
 2 & 0.9145 & 0.9077 \\ 
 3 & 0.9258 & 0.7925 \\ 
 4 & 0.9498 & -0.0359 \\ 
 \hline
 \hline
\end{tabular}
\caption{Regression scores of the Regression on the intermediate output in the case of the motion of a particle in a central potential with fixed angular momentum.}
\label{motionreg2}

\end{table}

\subsection{Electromagnetism}
\subsubsection*{Introduction}

Finally, we consider  electric $\mathbf{E}$ and magnetic fields $\mathbf{B}$, and their behaviour under Lorentz transformations. For this purpose we incorporate the fields in the electromagnetic field strength tensor

\begin{align}
    F_{\mu \nu}=\begin{pmatrix}
    0   & E_x& E_y & E_z\\
    -E_x & 0  & -B_z  & B_y  \\
    -E_y & B_z  & 0  &  -B_x \\
    -E_z & -B_y  & B_x  & 0  
    \end{pmatrix} \quad .
\end{align}
The Lorentz transformation of the field strength tensor 
\begin{align}
    F_{\mu \nu}^\prime= F_{\alpha \beta} \Lambda_{\ \mu}^\alpha \Lambda_{\ \nu}^\beta
\end{align}
implies the transformations for the electric and magnetic fields. The known Lorentz invariants of the electromagnetic fields are the determinant of the field strength tensor $\mathbf{B} \cdot  \mathbf{E}=\det{F}$ and $ \lvert\mathbf{B}\rvert^2- \lvert\mathbf{E}\rvert^2=1/2 F_{\mu\nu} F^{\mu \nu} $.

\subsubsection*{SNN Training}

\begin{center}
\begin{figure}[htb!]
\includegraphics[width=0.5\textwidth]{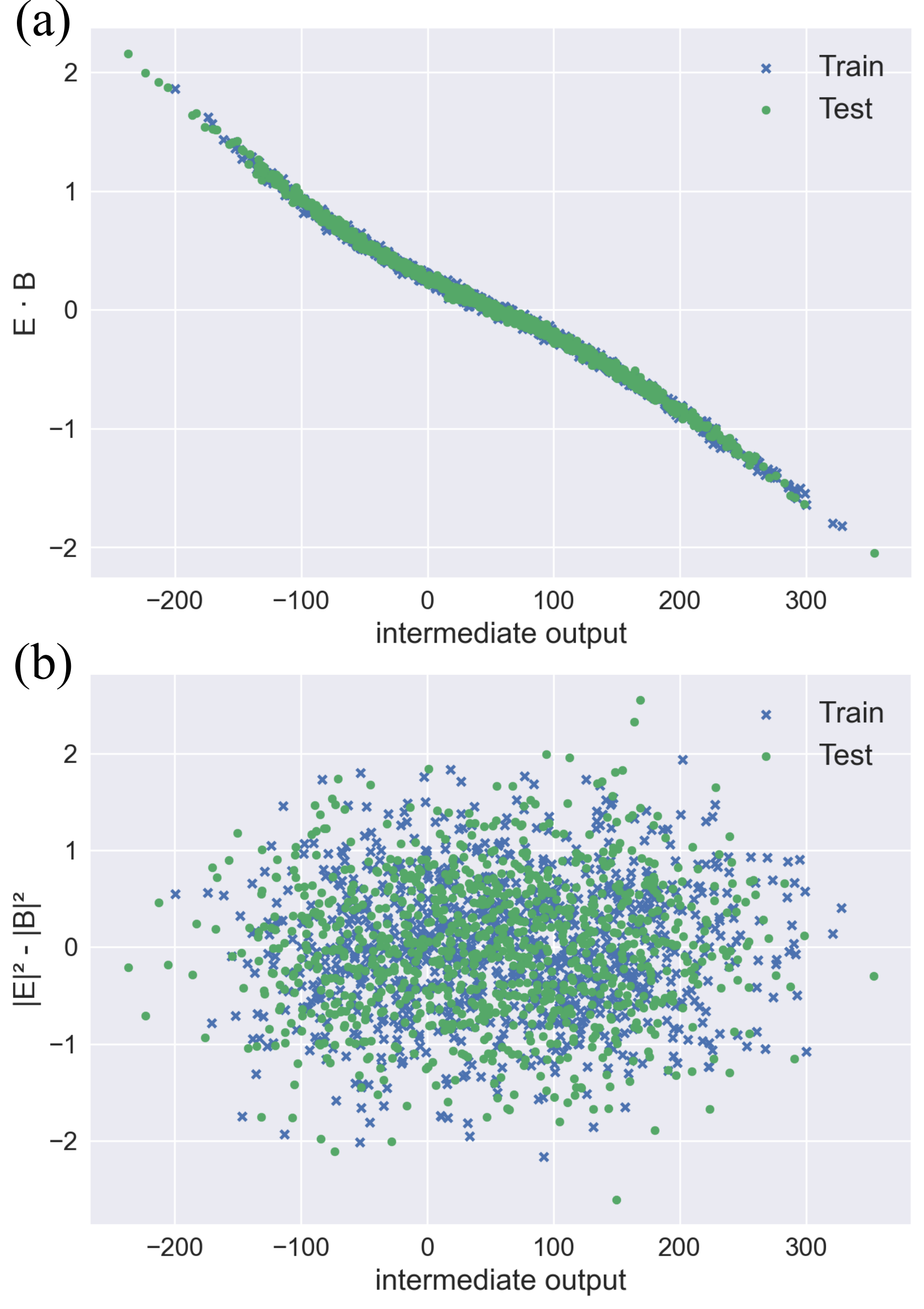}\\
\caption{Electromagnetism: Correlation between the intermediate output and (a) the determinant of the field strength tensor or (a) a specific contraction of two field strength tensors.}
\label{fig:edotb}
\end{figure}
\end{center}
\begin{center}
\begin{figure}[htb!]
\includegraphics[width=0.5\textwidth]{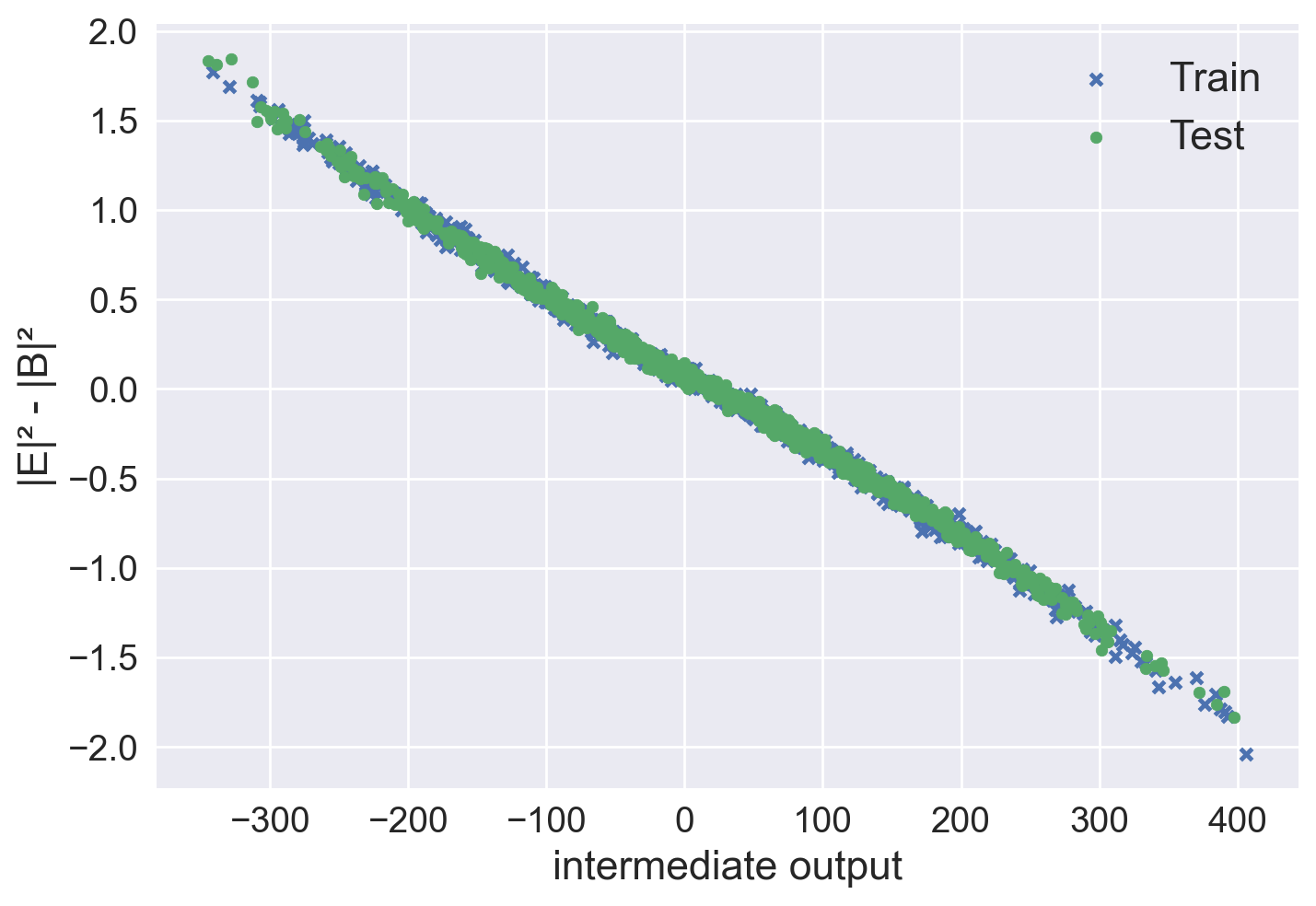}\\
\caption{Electromagnetism with fixed determinant of the field strength tensor: Correlation between the intermediate output and a specific contraction of two field strength tensors.}\label{fig:eembb}
\end{figure}
\end{center}

In this section, we study the behaviour of electromagnetic fields under Lorentz transformations with SNNs. For this purpose, we again produce 200000 true pairs $X=((E_x,E_y,E_z,B_x,B_y,B_z),(E_x^\prime,E_y^\prime,E_z^\prime,B_x^\prime,B_y^\prime,B_z^\prime))$ of electric and magnetic field configurations which are connected by a Lorentz transformation, and 200000 negative pairs of fields by permuting the positive pairs. 

We again train the SNN to predict if the two measurements belong to the same field configuration. After having successfully trained the neural network, we find that the neural network can fulfill the task to the high accuracy of $\approx 95\%$ on the training set and $\approx 94\%$ on the test set.

In order to determine what the neural network has learned, we perform polynomial regression on the intermediate output of the neural network. The function which approximates the output best is of degree two (see \tab{EMreg1}) and is given by

\begin{table}
\begin{tabular}{ c c c } 
 \hline
 \hline
 \hphantom{o} order \hphantom{o} &\hphantom{o} train score \hphantom{o}& \hphantom{o}test score \hphantom{o}\\ 
 \hline
 1 & 0.0000 & 0.0000 \\ 
 2 & 0.9902 & 0.9902\\ 
 3 & 0.9902 & 0.9902 \\ 
 4 & 0.9946 & 0.9946 \\ 
 \hline
 \hline
\end{tabular}
\caption{Regression scores of the Regression on the intermediate output in the case of Electromagnetism.}
\label{EMreg1}
\end{table}

\begin{align}
f(\mathbf{x})\approx& -170.53 E_2 B_2 -170.22 E_1 B_1  -170.20 E_3 B_3 \nonumber\\
& -4.13B_3^2+...+4.92 E_2^2 + 53.43 \nonumber \\
\approx & -170 \underbrace{(E_1 B_1 + E_2 B_2 + E_3 B_3)}_{=E \cdot B}+ 53
\end{align}

This function is an approximation to a known invariant, the determinant of the field strength tensor $\mathbf{B} \cdot  \mathbf{E}=\det{F}$. A confirmation of this deduction is the correlation between $\det{F}$ and the intermediate output as depicted in \fig{fig:edotb}.

Let us perform the same experiment again, however this time we fix the determinant of the field strength tensor when sampling the pairs of electromagnetic field configurations. The neural network still trains successfully and performs well in identifying pairs of data belonging to the same fields, with an accuracy of $\approx 93\%$ on the training set an $\approx 92\% $ on the test set. Performing the bottleneck regression on the intermediate output of the neural network reveals the remaining invariant to be of degree two (see \tab{EMreg2}) and is approximated by
\begin{table}
\begin{tabular}{ c c c } 
 \hline
 \hline
 \hphantom{o} order \hphantom{o} &\hphantom{o} train score \hphantom{o}& \hphantom{o}test score \hphantom{o}\\ 
 \hline
 1 & 0.0002 &  -0.0003 \\ 
 2 & 0.9956 & 0.9956\\ 
 3 & 0.9957 & 0.9956 \\ 
 4 &  0.9962 &  0.9962 \\ 
 \hline
 \hline
\end{tabular}
\caption{Regression scores of the Regression on the intermediate output in the case of Electromagnetism with fixed determinant of the field strength tensor.}
\label{EMreg2}
\end{table}

\begin{align}
f(\mathbf{x})\approx& -216.26 E_2^2  -216.016 E_1^2  -215.59 E_3^2 \nonumber \\
&-1.83 E_1 B_2 +...+  5.55 E_3 B_3+ 13.59\nonumber\\
&+215.80 B_3^2 + 216.57 B_2^2+  217.31 B_1^2\nonumber\\
\approx & -216\underbrace{(E_1^2+E_2^2+E_3^2-B_1^2-B_2^2-B_3^2)}_{=|E|^2-|B|^2}+14\quad .
\end{align}

This function is another known invariant of the field strength tensor $ \lvert\mathbf{B}\rvert^2- \lvert\mathbf{E}\rvert^2=1/2 F_{\mu\nu} F^{\mu \nu} $, confirmed in \fig{fig:eembb}. To summarize, in the context of electromagnetism, we have revealed the two invariants of the electric and magnetic fields which are preserved under Lorentz transformations.

\section{Conclusions and Future Directions}

We have introduced siamese neural networks (SNNs) to the field of theoretical physics. They are successful in predicting whether two data instances are connected by a deterministic transformation. We examined spacetime events and electromagnetic fields which transform under Lorentz transformations, as well as the movement of particles in a central potential. By interpreting our neural network, we found that it learns the underlying symmetry invariants and conserved quantities to perform its prediction. Most interestingly, we were able to interpret our SNNs via the use of polynomial regression. This procedure revealed an excellent approximation of the underlying symmetry invariants and conserved quantities. These invariants range from the spacetime interval over angular momentum conservation to the determinant of the field strength tensor. If the underlying system does not contain human readable invariants, the neural network could act as an approximation to such an invariant.

Future directions of this work include an upgrade of the polynomial regression to symbolic regression \cite{Koza1994}. Another exciting direction is to combine interpretable SNNs with semi-automated mathematical reasoning tools, e.g., solvers or theorem provers. The idea is to check the physical law learned by the SNN for consistency against known laws and invariants by leveraging such reasoning tools \cite{lgml}. It does not take much imagination to envision how this technology can be used in applications such as quantum error correction or in particle tracking at the LHC. 

It remains to be seen if SNNs will ever find an invariant or conserved quantity unknown to modern physics. Even if this does not happen, the contribution of this work is the introduction of SNNs as a useful tool in theoretical physics. Furthermore, we challenged the black box nature of artificial neural networks by a very clear interpretation that reveals polynomial quantities without prior knowledge. The interpretation procedure might also be adopted into the field of computer science, where the interpretability of neural networks poses a major problem.


\section{Acknowledgements}
We thank Isaac Tamblyn for helpful discussions. RGM and JS are supported by NSERC. RGM is further supported by the Canada Research Chair program, and the Perimeter Institute for Theoretical Physics. We thank the National Research Council of Canada for their partnership with Perimeter on the PIQuIL. Research at Perimeter Institute is supported in part by the Government of Canada through the Department of Innovation, Science and Economic Development Canada and by the Province of Ontario through the Ministry of Colleges and Universities. 

 

\appendix
\section{Lorentz Transformation}
\label{app:lt}
Let us describe the representation of the Lorentz transformations which are used to generate the data pairs in the special relativity and electromagnetism sections.

An arbitrary Lorentz transformation can be decomposed

\begin{align}
    \Lambda = D_1 \Lambda_v D_2
\end{align}

Here $\Lambda_v$ is a Lorentz boost in $x$ direction.

\begin{align}
    \Lambda_v=\begin{pmatrix}
    \gamma   & -\gamma \beta & 0 & 0\\
    -\gamma\beta &  \gamma & 0  & 0  \\
    0 & 0  & 1  &  0 \\
    0 & 0  & 0  & 1  
    \end{pmatrix}
\end{align}

where

\begin{align}
    \beta=\frac{v}{c}\quad , \quad \gamma=\frac{1}{\sqrt{(1-\beta^2)}} \, .
\end{align}

$c$ is the speed of light which we conveniently set to $c=1$.

The matrices $D_1,D_2$ perform the rotation in the three dimensional subspace

\begin{align}
    D=\begin{pmatrix}
    1   & 0\\
    0&  \mathcal{R} 
    \end{pmatrix}
\end{align}

where $\mathcal{R}\in \mathrm{O}(3)$

\section{Neural Network Details}
\label{app:nn}

\begin{center}
\begin{figure}[h!]
\includegraphics[width=0.5\textwidth]{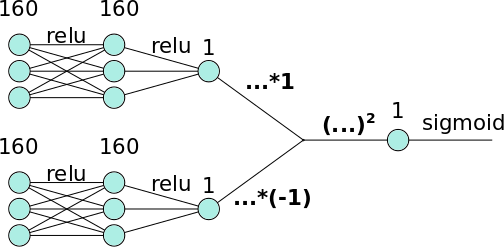}\\
\caption{Detailed architecture of an interpretable siamese neural network.}
\label{fig:neuronnumber01}
\end{figure}
\end{center}

In this section we explain the details of the training of the SNN on pairs of data with a number of datapoints $N$ between 50000 an 200000. For the sake of understandability we use the same architecture and hyperparameters for all learning tasks. The architecture of the SNN is depicted in \fig{fig:neuronnumber01}.

\begin{center}
\begin{figure*}[htb!]
\includegraphics[width=0.7\textwidth]{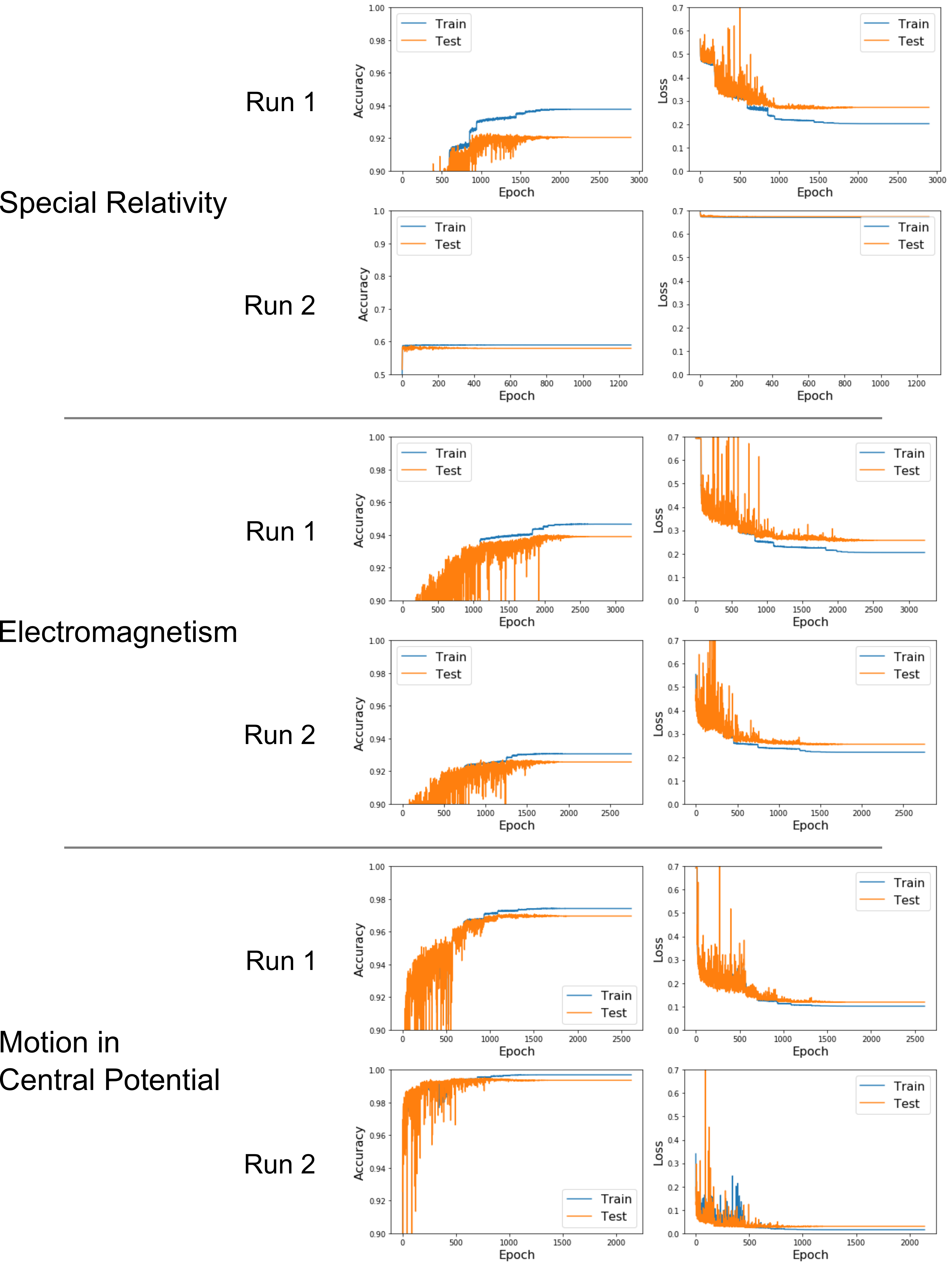}\\
\caption{Losses and Accuracies}
\label{fig:lossacc}
\end{figure*}
\end{center}

The training of the neural network is the adjustment of the weights $w_{ij}^L$ and biases $b_i^L$ of the neural network to achieve a minimum of the binary cross entropy loss function for all $N$ training datapoints. 
\begin{align}
    L(y_{t},y_{p})=  -\frac{1}{N}\sum_{i=0}^N y_{i,t} \ln (y_{i,p})+(1-y_{i,t})\ln (1-y_{i,p}) \quad ,
\end{align}
where $y_t$ denotes the true label, while $y_p$ is the neural network prediction. Our neural networks are trained using the Adadelta optimizer. We found that starting learning rates of $lr = 100$ are needed to train the neural network, this learning rate is a lot higher than normally used in traditional classification problems. Each  update is performed by calculating the gradient on a batch of size 256. We employ learning rate decay callbacks which reduce the learning rate by a factor of 2 if the training loss has not improved for 50 epochs. We train our networks for 10000 epochs however, we employ an early stopping callback which aborts the training process if the training loss has not improved over 200 epochs. We do not use any kind of regularization in our neural networks. The evolution of the losses and accuracies during training are depicted in \fig{fig:lossacc}

\section{A Note on Interpreting Neural Networks}
\label{app:note}
If a neural network intrinsically learns a physical observable $\mathcal{O}(x)$ as a function of the input data $x$, this observable is often encoded in an elusive manner distributed among many neurons. The bottleneck interpretation forces all information of this observable through a single neuron. In general, this observable is encoded in a deformed manner such that the output of the bottleneck neuron is $h(\mathcal{O}(x)))$. If we restrict ourselves to a small output range, the function $h$ can be linearized such that $h(\mathcal{O}(x)))=h_0+h_1\times\mathcal{O}(x)$. This form helps us to perform linear regression and $h_0$ and $h_1$ can be identified as adjustments to the weights and bias of the neuron.

\bibliographystyle{apsrev4-1}
\bibliography{SiameseBib}


\end{document}